\documentclass[10pt]{article}
\usepackage{amsmath}
\usepackage{amssymb}
\usepackage{graphicx}

\def\aa{A\&A}

\def\fracd#1#2{{\displaystyle\frac{#1}{#2}}}

\begin{document}

\setcounter{figure}{0}
\setcounter{table}{0}
\setcounter{footnote}{0}
\setcounter{equation}{0}

\vspace*{0.5cm}

\noindent {\Large \strut ON ERRORS OF RADIO SOURCE POSITION CATALOGS}
\vspace*{0.7cm}

\noindent\hspace*{1.5cm} Z. MALKIN\\
\noindent\hspace*{1.5cm} Pulkovo Observatory, St. Petersburg, Russia\\
\noindent\hspace*{1.5cm} St. Petersburg State University, St. Petersburg, Russia\\
\noindent\hspace*{1.5cm} e-mail: malkin@gao.spb.ru\\

\vspace*{0.5cm}

\noindent {\large ABSTRACT.}
In this paper, a new method of investigation of the external radio source position catalogs RSPCs stochastic errors is presented.
Using this method the stochastic errors of nine recently published RSPCs were evaluated.
It has been shown that the result can be affected by the systematic differences between catalogs if the latter are not accounted for.
It was also found that the formal uncertainties of the source position in the RSPCs correlate with the external errors.
We also investigated several topics related to the formal uncertainties and systematic errors of RSPC.

\vspace*{1cm}

\noindent {\large 1. INTRODUCTION}

\smallskip

VLBI is currently the primary technique for maintening International Celestial Reference Frame (ICRF, Ma et al. 2009).
The latter is realized as a catalog of radio source coordinates derived from processing of VLBI observations.
Assessing the systematic and stochastic errors of radio source position catalogs (RSPCs) plays an important role in improvement of the ICRF.
The internal stochastic error of the RSPCs is determined by the source position uncertainties given in the catalog.
The external stochastic error can be assessed only from mutual comparison of several RSPCs.

In this work, we present a new approach to computation of the external stochastic errors of RSPCs.
It allows to simultaneously analyze an unlimited number of RSPCs, the more the better, in fact.
A key point is a new method of estimation of the correlation between catalogs.
Another development is a new concept of weighted correlation coefficient, which is important for analysis of unevenly weighted data.
The third improvement is accounting for systematic differences between catalogs.
With this method, we obtained errors of nine recently published RSPCs.
See Malkin (2013) for detailed description of the method and results.

We also investigated several other topics related to the formal uncertainties of the ICRF2 sources and a correspondence between 
the formal and external errors in source position.

%%%%%%%%%%%%%%%%%%%%%%%%%%%%%%%%%%%%%%%%%%%%%%%%%%%%%%%%%%%%%%%%%%%%%%%%%%%%%%%%%%%%%%%%%%%%%%

\vspace*{0.7cm}

\noindent {\large 2. METHOD OF THE ASSESSMENT OF EXTERNAL CATALOG ERRORS}

\smallskip

We base our analysis at the 3-cornered hat method (TCH) originally developed for investigation of the clock frequency instability.
In its original formulation, the TCH method was applied to three series of measurements, however it can be generalized to N-cornered-hat (NCH) method.
If we analyze N catalogs, we have to solve the following system:

\begin{equation}
\sigma^2_{ij} = \sigma^2_i + \sigma^2_j -2 \rho_{ij} \sigma_i \sigma_j, \quad i=1 \ldots N-1, \ j=i+1 \ldots N \,, \\
\label{eq:nch_corr_equation}
\end{equation}

\noindent where $\sigma_{ij}$ are variances of paired differences between catalogs, $\rho_{ij}$ are correlation coefficients between catalogs, 
$\sigma_i$, $\sigma_j$ are unknown external errors of catalogs.
For $N$ catalogs, we have $N(N-1)$ equations.

The key point of the method is to find reliable estimates of the correlation coefficients $\rho_{ij}$.
We propose the following strategy to estimate $\rho_{ij}$.
Let us have $N$ catalogs.
First we select sources in common in all the catalogs, which are used for the analysis.

Now we consider the $i$-th and $j$-th catalogs.
At the first step we computed the differences between these catalogs with all $k$-th catalogs, $k=1,...,N, \ k \neq i, \ k \neq j$.
After that, we computed the correlation $\rho^k_{ij}$ between catalog differences $\Delta_{ik} = Cat_i - Cat_k$
and $\Delta_{jk} = Cat_j - Cat_k$ for each $k$, where $Cat_i, Cat_j$, and $Cat_k$ are vectors of the source positions in common.
Computations were made separately for right ascension (RA) and declination (DE).
RA differences were multiplied by $\cos (DE)$.
The average value of $\rho^k_{ij}$ over all $k$ was considered an approximation to the correlation $\rho_{ij}$ between $i$-th and $k$-th catalogs.

\begin{figure}[ht!]
\begin{minipage}{0.52\hsize}
  To estimate the correlation coefficient between two RSPCs we used a weighted correlation coefficient defined as
  \begin{equation}
  \rho^{w}_{xy} = \fracd{\sum\limits_{i} \sqrt{p_{x,i}p_{y,i}}(x_i-\bar{x})(y_i-\bar{y})}{\sqrt{\sum\limits_{i} p_{x,i}(x_i-\bar{x})^2\sum\limits_{i} p_{y,i}(y_i-\bar{y})^2}} \,,
  \label{eq:wcorr}
  \end{equation}
  \noindent where $x_i$ and $y_i$ are input data, $s_{x,i}$ and $s_{y,i}$ are their standard errors, 
  $p_{x,i} = 1/s_{x,i}^2 , p_{y,i} = 1/s_{y,i}^2$, $\bar{x}$ and $\bar{y}$ are weighted mean of $x_i$ and $y_i$.
  Figure \ref{fig:corr} shows an example of computation of the standard ($\rho_{xy}$) and weighted ($\rho^{w}_{xy}$) correlation coefficient for an
  artificial set consisting of five measurements with two outliers.
  In this example, the standard correlation coefficient is equal to zero, whereas the weighted correlation coefficient is about unity.
\end{minipage}
\hfill
\begin{minipage}{0.42\hsize}
  \centering
  \includegraphics[clip,width=\hsize]{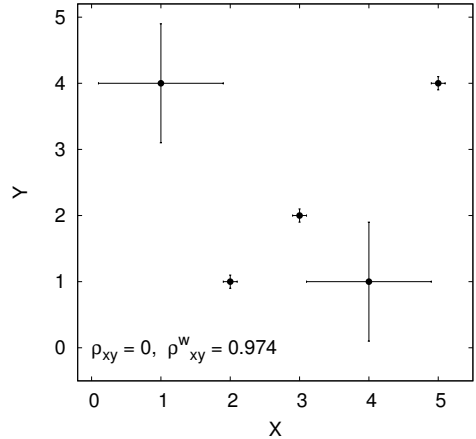}
  \caption{Standard ($\rho$) and weighted ($\rho^{w}$) correlation coefficient.}
  \label{fig:corr}
\end{minipage}
\end{figure}

%%%%%%%%%%%%%%%%%%%%%%%%%%%%%%%%%%%%%%%%%%%%%%%%%%%%%%%%%%%%%%%%%%%%%%%%%%%%%%%%%%%%%%%%%%%%%%

\vspace*{0.7cm}

\noindent {\large 4. RESULTS OF ANALYSIS}

\smallskip

We investigated nine recently published RSPCs: aus2012b, bkg2012a, cgs2012a, gsf2012a, igg2012b, opa2013a, rfc2013a, sha2012b, and usn2012a.
They have 703 sources in common, which were used in subsequent computations.

\begin{figure}[th]
\centering
\includegraphics[clip,width=\hsize]{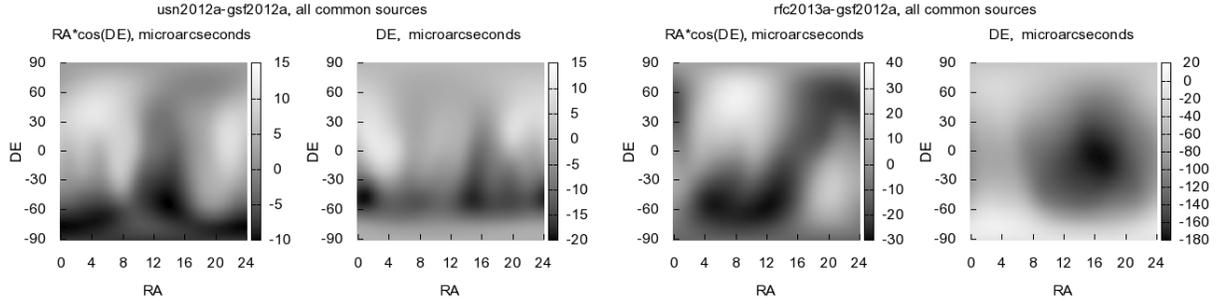}
\caption{Systematic differences between catalogs: two examples of small (on the left) and large (on the right) differences. Unit: $\mu$as.
See (Malkin 2013) for more plots.}
\label{fig:systematics}
\end{figure}

The systematic differences between catalogs may have a substantial impact on the determination of their stochastic errors.
Two examples of the systematic differences are depicted in Fig~\ref{fig:systematics}.
Note larger differences in declination as compared with the differences in right ascension.
One can see that the systematic differences between catalogs have a complicated structure, which cannot be described in terms of rotation
and a few supplement low-frequency terms. 

The variances of the paired differences and correlation coefficients between catalogs were computed both for the original differences and the
differences corrected for the systematic differences.
It was found that both the variances of paired differences and the correlation coefficients between catalogs are substantially affected by
the systematics (Malkin 2013) .
The effect is especially significant for the pairs of catalogs with large systematic differences.

\begin{figure}[ht!]
\centering
\includegraphics[clip,width=0.9\hsize]{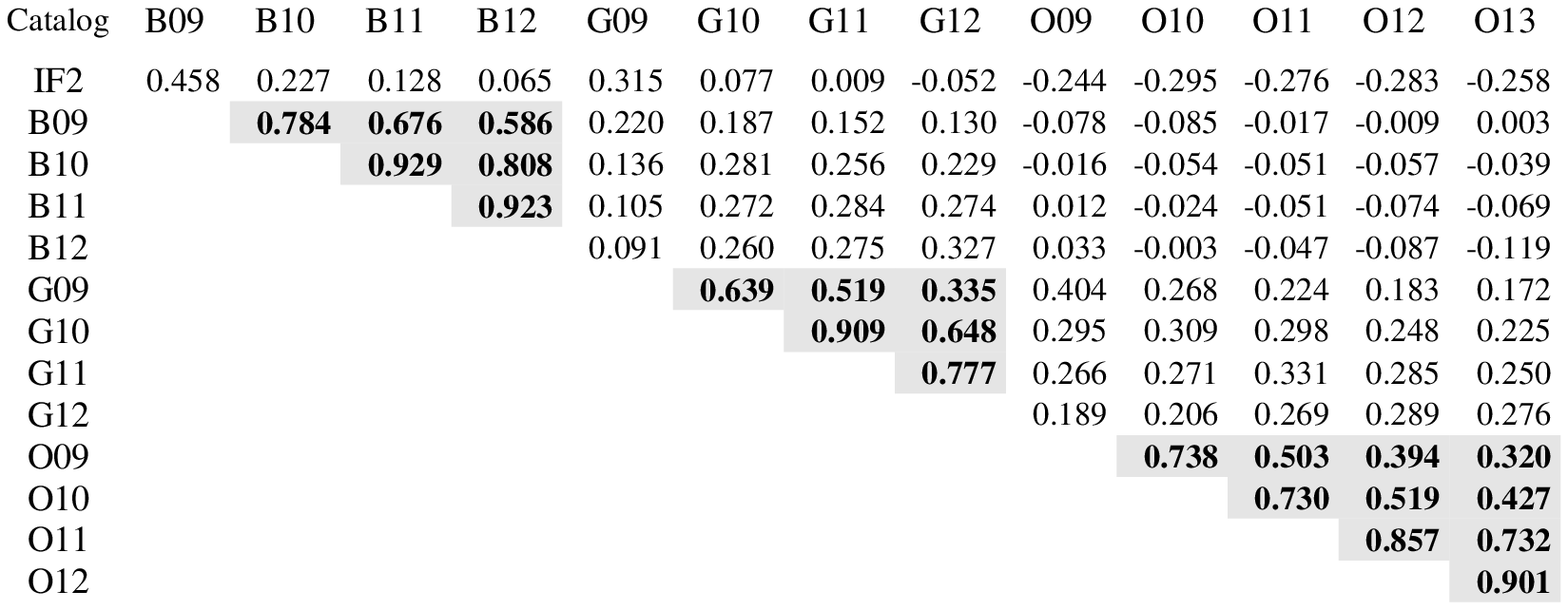}
\caption{Correlation coefficients between input catalogs. The catalog designation is formed from the first letter of the AC name (B for BKG, G for GSFC,
  O for OPA) and two last digits of year of publication; IF2 stands for ICRF2. The values related to catalogs obtained at the same AC are highlighted.}
\label{fig:catcorr}
\end{figure}

Table~\ref{tab:corr} shows the standard and weighted correlation coefficient between GSF and other catalogs.
One can see that the correlations in RA and DE are very similar, and there is no clear dependence on the software.
It is also noticeable that catalogs obtained at the same AC more closely correlate with each other than catalogs obtained in different ACs (Fig~\ref{fig:catcorr}).

\begin{table*}
\centering
\begin{tabular}{ccccccc}
\hline
Catalogs & $\rho$ & $\rho^w$  \\
         & $\alpha$ / $\delta$ & $\alpha$ / $\delta$ \\
\hline
AUS -- GSF & $ +0.1861 \ / +0.2032 $ & $ +0.1125 \ / +0.1129 $ \\
BKG -- GSF & $ +0.5082 \ / +0.6095 $ & $ +0.4794 \ / +0.5038 $ \\
CGS -- GSF & $ +0.7711 \ / +0.7746 $ & $ +0.6395 \ / +0.6348 $ \\
GSF -- IGG & $ -0.0193 \ / +0.2334 $ & $ +0.3732 \ / +0.3693 $ \\
GSF -- OPA & $ +0.5210 \ / +0.4823 $ & $ +0.4711 \ / +0.4931 $ \\
GSF -- RFC & $ -0.2497 \ / -0.0311 $ & $ +0.2184 \ / +0.2582 $ \\
GSF -- SHA & $ +0.1528 \ / +0.1270 $ & $ +0.1218 \ / +0.1195 $ \\
GSF -- USN & $ -0.0268 \ / -0.0832 $ & $ +0.0099 \ / +0.0346 $ \\
\hline
\end{tabular}
\caption{Standard ($\rho$) and weighted ($\rho^w$) correlation coefficients between GSF and other catalogs.}
\label{tab:corr}
\end{table*}

\begin{figure}[ht!]
\begin{minipage}[b]{0.46\hsize}
  \centering
  \includegraphics[clip,width=\hsize]{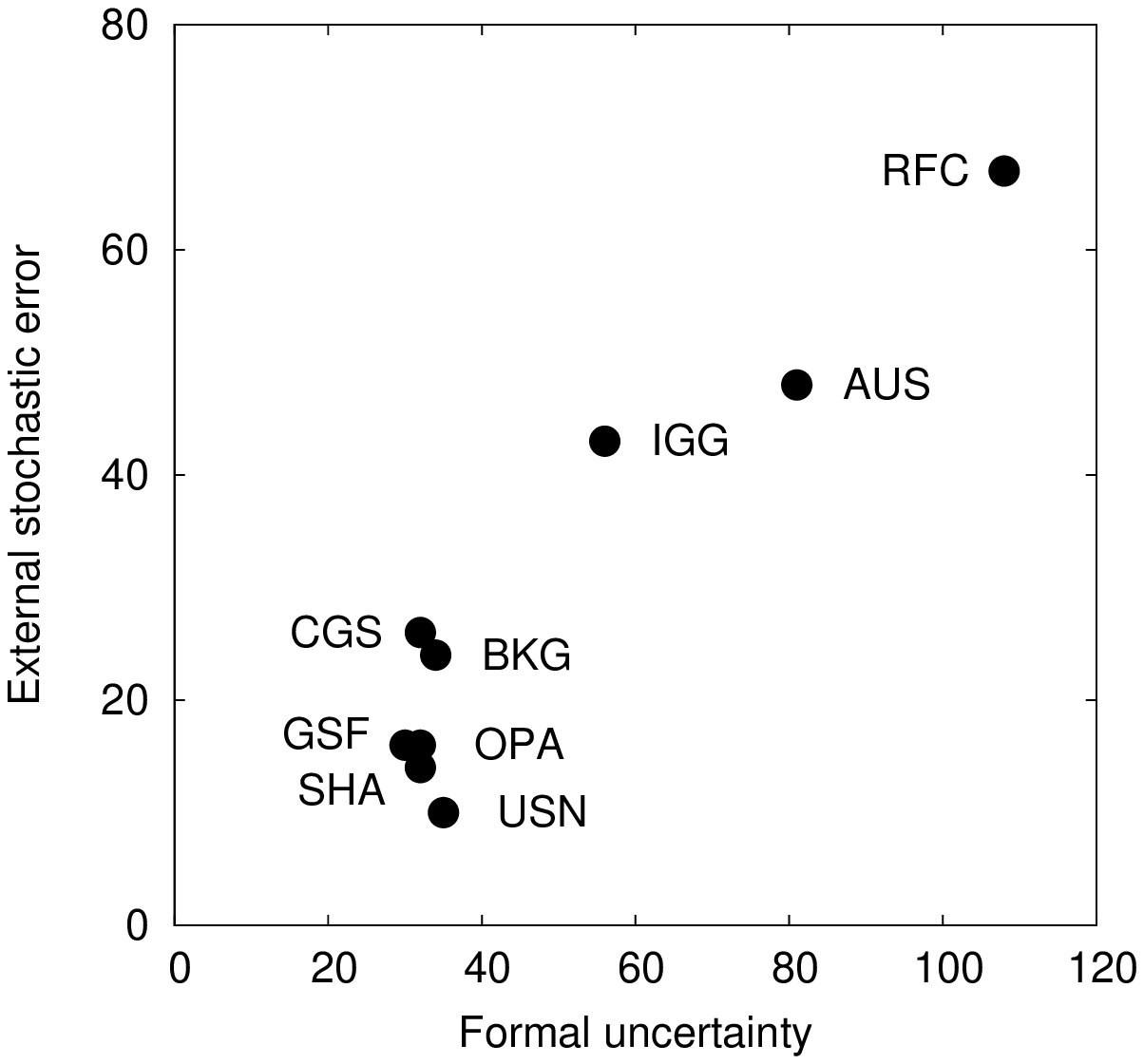}
  \caption{Correlation between reported and external errors.}
  \label{fig:errors}
\end{minipage}
\hfill
\begin{minipage}[b]{0.49\hsize}
  \centering
  \includegraphics[clip,width=\textwidth]{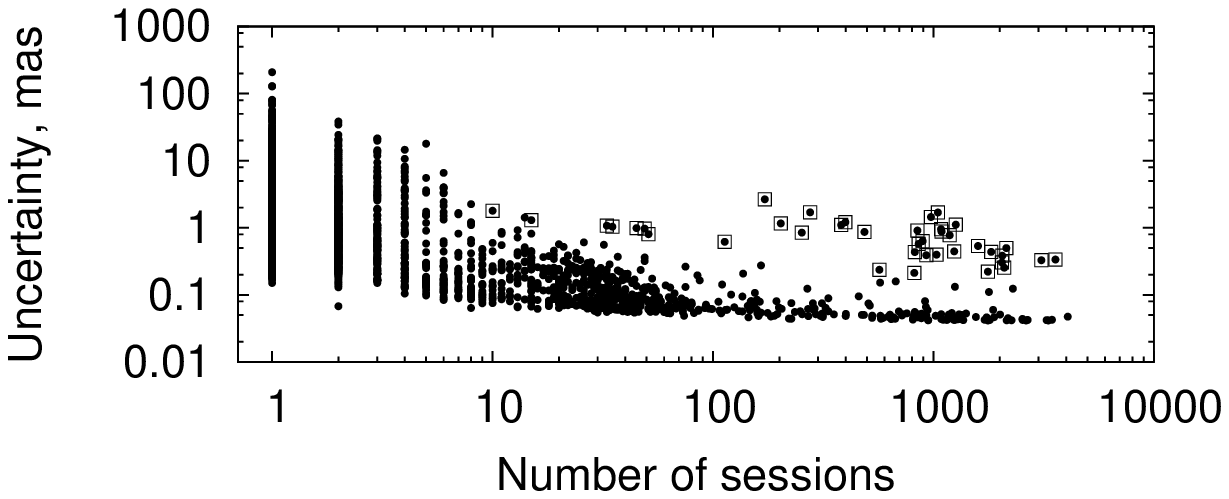}
  \includegraphics[clip,width=\textwidth]{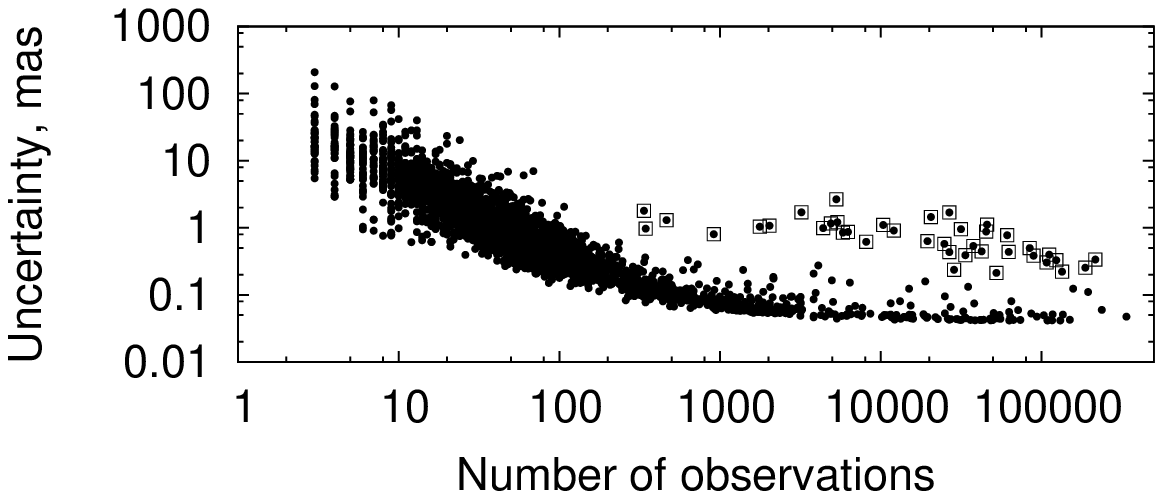}
  \caption{Dependence of the source position uncertainty on the observational history.}
  \label{fig:gsfc_errors}
\end{minipage}
\end{figure}

Finally, we computed the stochastic errors of the nine RSPCs in two ways: with and without correcting for the systematic differences between catalogs.
The weighted correlation coefficients were used in both cases.
The results are presented in Table~\ref{tab:errors}.
A comparison of the two variants shows that the systematic differences significantly affect the determination of their stochastic accuracy.
The numbers in the last column of Table~\ref{tab:errors} are considered as the final result of our work.
For comparison, the median source position uncertainty as reported in the catalog is given in the second column of Table~\ref{tab:errors}.
Figure~\ref{fig:errors} shows that the external stochastic errors correlate with formal (reported) uncertainties for catalogs with relatively
large errors.
For catalogs with higher accuracy, such a dependence is mush smaller. 

\begin{table}
\centering
\begin{tabular}{cccc}
\hline
Catalog &    ME  & Original differences & Corrected differences \\
          &$\alpha$ / $\delta$& $\alpha$ / $\delta$ &  $\alpha$ / $\delta$ \\
\hline                          
AUS       & ~76 / ~86       & 49 / 56 & 46 / 51 \\
BKG       & ~28 / ~40       & 23 / 27 & 21 / 27 \\
CGS       & ~26 / ~38       & 27 / 46 & 25 / 27 \\
GSF       & ~24 / ~36       & 15 / 21 & 14 / 17 \\
IGG       & ~49 / ~62       & 48 / 59 & 42 / 44 \\
OPA       & ~27 / ~37       & 15 / 23 & 14 / 18 \\
RFC       & 105 / 110       & 63 / 93 & 60 / 74 \\
SHA       & ~27 / ~38       & 13 / 17 & 12 / 17 \\
USN       & ~29 / ~41       & 10 / 12 & 10 / 10 \\
\hline
\end{tabular}\\
\caption{Median reported uncertainties (ME) and external stochastic errors computed using original differences and differences corrected for the systematics. Unit: $\mu$as}
\label{tab:errors}
\end{table}

Figure~\ref{fig:gsfc_errors} shows the dependence of the position uncertainty in the ICRF2 catalog 2012a on the number of sessions
and the number of observations (delays).
Although the number of delays closely correlates with the number of sessions, the former seems to be the better argument for description
of the dependence of the position uncertainty on the observational history of the source.

During computation of ICRF2, positions and position uncertainties of 39 sources were not solved as global parameters like positions and position
errors of other {\it global} sources, but was derived from a special analysis of source position time series (Ma et al. 2009).
For this reason, these sources were referred to as special handling sources (a.k.a. {\it arc} sources).
As one can see in Fig.~\ref{fig:gsfc_errors}, formal errors in position of {\it arc} sources (marked with squares) do not correspond to general law.
This problem was earlier addressed in the IERS/IVS ICRF2 Working Group discussions, but has not been satisfactory solved until now.
Evidently, a special procedure to compute the position errors of these sources should be developed.
On the other hand, a necessity for including {\it arc} sources in the ICRF may be worth further discussion.

%%%%%%%%%%%%%%%%%%%%%%%%%%%%%%%%%%%%%%%%%%%%%%%%%%%%%%%%%%%%%%%%%%%%%%%%%%%%%%%%%%%%%%%%%%%%%%

\vspace*{0.7cm}

\noindent {\large 4. CONCLUSION}

\smallskip

1. A new approach to assess the external stochastic errors of radio source position catalogs has been developed.
The new features of this method are: simultaneous processing of all catalogs, implementing a new strategy for estimating the correlations between RSPCs,
using weighted correlation coefficients, accounting for systematic differences between RSPCs.
Using this approach, we obtained independent estimates of the stochastic errors of the nine recently published catalogs, some of them for the first time.

2. Modern radio source position catalogs show significant and complicated systematic differences at a level of tens $\mu$as, which must be accounted
for during accuracy assessment and combination.

3. Catalogs obtained at the same AC are in close correlation with each other.
This may evidence the presence of AC-specific systematic errors caused by specific modeling and analysis options traditionally used at different ACs.

4. The external catalog stochastic errors closely correlate with the formal source position uncertainty, most probably because of quality
of the software used and analysis strategy details such as modelling and parameterization.

5. The ICRF2 source position uncertainties are not homogeneous for {\it global} and {\it arc} sources, which should be addressed during preparation
of next ICRF versions.

\bigskip
\noindent {\it Acknowledgements.}
The author is grateful to all the authors of the RSPCs, who made them available to us either via public access (AUS, BKG, CGS, GSF, OPA, RFC)
or via personal contact (IGG, SHA, USN).
The travel support from the organizers of the conference is highly appreciated.

\vspace*{0.7cm}

\noindent {\large 4. REFERENCES}

{

\leftskip=5mm
\parindent=-5mm

\smallskip

Ma, C., Arias, E.F., Bianko, G., et al., 2009, ``The Second Realization of the International Celestial Reference Frame by Very Long Baseline Interferometry'',
IERS Technical Note No.~35, A.~Fey, D.~Gordon, C.S.~Jacobs (eds.). Frankfurt am Main: Verlag des Bundesamts f\"ur Kartographie und Geod\"asie.
%Ma, C., Arias, E., Bianko, G., et al., 2009, IERS Technical Note 35, A.Fey, D.Gordon, C.S.Jacobs (eds.).

Malkin, Z., 2013, ``A new approach to the assessment of stochastic errors of radio source position catalogues''. \aa, 558, A29.

}

\end{document}